\pdfoutput=1
\documentclass[a4paper, 12pt]{article}
\usepackage[title]{appendix}    

\usepackage[T1]{fontenc}
\usepackage[utf8]{inputenc}
\usepackage{lmodern}
\usepackage{textcomp}
\DeclareUnicodeCharacter{2009}{\,}
\usepackage[english]{babel}
\usepackage{csquotes} 
\usepackage{longtable}
\usepackage{array}
\usepackage{booktabs} 

\usepackage{amsmath, amssymb, amsfonts}
\usepackage{bm}

\usepackage[margin=1in]{geometry}
\usepackage{graphicx}
\graphicspath{{figs/}}

\usepackage{makecell} 
\usepackage{multirow}
\usepackage{float}
\usepackage{subcaption}
\usepackage{booktabs}

\usepackage[dvipsnames]{xcolor} 
\definecolor{linkcolor}{rgb}{0.0,0.3,0.5}
\usepackage[hypertexnames=false, unicode, colorlinks=true,
linkcolor=linkcolor, citecolor=linkcolor,
filecolor=linkcolor, urlcolor=linkcolor,
pdfusetitle]{hyperref}
\usepackage[all]{hypcap}

\usepackage{authblk}
\usepackage{verbatim}
\usepackage{xspace}
\usepackage{microtype}
\usepackage[none]{hyphenat}
\usepackage{cleveref}
\DeclareMathAlphabet{\mathpzc}{OT1}{pzc}{m}{it}

\makeatletter
\def\@seccntformat#1{\hspace*{0pt}\csname the#1\endcsname.\quad}
\def\section{\@startsection{section}{1}{0pt}{3.5ex plus 1ex minus .2ex}{2.3ex plus .2ex}{\centering\normalfont\large\bfseries}}
\makeatother

\title{\textbf{\boldmath Constraining ADD black holes at the LHC with $\sqrt{s} = 14$ TeV}}
\author[]{Ashfaque Ahmad\thanks{gq0084@myamu.ac.in}}
\author[]{Sudhir Kumar Gupta\thanks{sudhir.ph@amu.ac.in}}
\author[]{Abbas Ali\thanks{pht05aa@gmail.com}}

\affil[]{Department of Physics, Aligarh Muslim University, Aligarh, UP -- 202002, India}

\date{} 

\begin{document}
	\maketitle
	\begin{abstract}
		We explore the possibility of the microscopic black hole formation at the Large Hadron Collider (LHC) in the context of the ADD model, for the centre-of-mass energy $\sqrt{s} = 14~\mathrm{TeV}$ with an integrated luminosity of $349.4~\mathrm{fb}^{-1}$ and provide constraints on the black hole mass, $M_{\mathrm{B}}$ by taking into account the effects of loss during the formation process of black holes through the parameter $\zeta$. Our analysis reveals that for $\zeta = 0$, black holes with $M_{\mathrm{B}} \leq 11.83~\mathrm{TeV}$ are disfavored in the case of three extra dimensions ($\mathcal{D}$), for the reduced Planck scale ($\Lambda_{\mathcal{D}}$) of about a few TeV. The corresponding values for $\Lambda_{\mathcal{D}} = 9~\mathrm{TeV}$ turn out to be about $10.33~\mathrm{TeV}$. A significant reduction in the aforementioned limits is observed while the loss gets higher, e.g. for $\zeta = 0.35$, $M_{\mathrm{B}}$ reduces to $7.65~(6.82)~\mathrm{TeV}$ at $\Lambda_{\mathcal{D}} = 1~(9)~\mathrm{TeV}$. These limits change to $12.03 
		~(10.88)~\mathrm{TeV}$ and $7.80~ (7.03)~\mathrm{TeV}$ respectively for $\zeta = 0~(0.35)$ and $\Lambda_{\mathcal{D}} = 1~ (9)~\mathrm{TeV}$ at 95\% CL in case $\mathcal{D}$ is raised to seven.
	\end{abstract}
	\clearpage
	
	\section{Introduction}
	The discovery of the Higgs boson~\cite{ATLAS:2012yve} at the LHC~\cite{CMS:2012qbp} represented a pivotal confirmation of the Standard Model (SM). However, besides its remarkable success in describing fundamental particles and their interactions, the SM remains an incomplete theory due to its incapability in addressing issues related to naturalness problem of the electroweak scale~\cite{Susskind:1978ms, Weinberg:1975gm}, neutrino mass~\cite{Mohapatra:1979ia} and the existence of the dark matter~\cite{Bertone:2004pz, Clowe:2006eq}, which appears to fill much of the universe. These, in addition to the lack of gravitational interaction within the SM provide a strong motivation to explore physics beyond the SM (BSM). Among the potential candidates of BSM theories, model with additional spatial dimensions such as ADD model~\cite{Antoniadis:1998ig, Arkani-Hamed:1998jmv} by Arkani-Hamed, Dimopoulos, and Dvali, turned out to be interesting possibilities as these, in addition to providing a satisfactory explanation to the naturalness problem by means of reducing the Planck scale in the TeV range, also provide opportunity to accommodate gravity which could be testified experimentally via its signatures in various experiments~\cite{ATLAS:2012hvw, CMS:2012vzw, CMS:2021ctt, D0:2003mxa}. In scenarios where the fundamental Planck scale is of the order of a few TeV, gravitational interactions can become significant at LHC energies, potentially enabling the production of black hole. The black hole formation is motivated by the hoop conjecture~\cite{Thorne:1972ji}, according to which an event horizon forms when sufficient energy is concentrated within a region smaller than the corresponding Schwarzschild radius ($r_s$). These black holes are different from astrophysical black holes and are generally studied within the framework of high-energy physics. They are expected to evaporate very quickly through Hawking radiation~\cite{Hawking:1975vcx}. The black hole produced at the LHC are expected to be extremely small~\cite{Cavaglia:2002si}. For example, if black hole mass is $M_{\mathrm{B}} = 2~\mathrm{TeV}$ for extra spatial dimension, $\mathcal{D} = 3$, and reduced Planck mass, $\Lambda_{\mathcal{D}} = 1$, the Schwarzschild radius is about $\sim 4.3 \times 10^{-19}~\mathrm{m}$. This shows that these black holes would be microscopic.

	To search for evidence of these black holes, the ATLAS and CMS collaborations~\cite{Mandrik:2018gud,CMS:2017boz, CMS:2018ozv} have performed several analyses at various collision energies, and have ruled out the possibility of microscopic black hole with masses below a certain limit~\cite{Savina:2011zz, Savina:2013eja, Savina:2015zda}. Using LHC Run 1 data at $\sqrt{s} = 7~\mathrm{TeV}$ with an integrated luminosity of $35~\text{pb}^{-1}$, the CMS collaborations rule out black hole masses up to $4.5$ ($3.5$)~$\mathrm{TeV}$, corresponding to a $\Lambda_{\mathcal{D}} = 2.5$ ($3.5$)~$\mathrm{TeV}$ for $\mathcal{D}=6$ \cite{CMS:2010oej}. After the searches at $\sqrt{s}=7~\mathrm{TeV}$, the exclusion limits were extended using higher collision energy of $\sqrt{s}=8~\mathrm{TeV}$ with an integrated luminosity of $12.1~\text{fb}^{-1}$. For $\mathcal{D}=6$, black hole masses up to $6.2$ ($5.7$)~$\mathrm{TeV}$ were excluded for $\Lambda_{\mathcal{D}}=1.5$ ($5$)~$\mathrm{TeV}$. Similarly, for $\mathcal{D}=2$, the excluded masses reached $5.9$ ($5.2$)~$\mathrm{TeV}$ for $\Lambda_{\mathcal{D}}=1.5$ ($4.5$)~$\mathrm{TeV}$~\cite{CMS:2013ndg}. Further searches were performed using the run 2 dataset at a higher center-of-mass (CM) energy of $\sqrt{s}=13~\mathrm{TeV}$ with an integrated luminosity $\int\mathcal{L}dt=138~\mathrm{fb}^{-1}$. These searches excluded black hole masses up to $11.4$ ($10.4$)~$\mathrm{TeV}$ for $\mathcal{D}=6$ and up to $11$ ($9.4$)~$\mathrm{TeV}$ for $\mathcal{D} = 2$, corresponding to $\Lambda_{\mathcal{D}} = 2$ ($8$)~$\mathrm{TeV}$ at $95\%$ C.L.~\cite{ CMS:2026cgd, ATLAS:2023vat, Vami:2025fsl, ATLAS:2026hzb}. In all the black hole searches discussed above, black holes are assumed to be form in the semi-classical regime, where their masses $M_{\mathrm{B}}$ are greater than the $\Lambda_{\mathcal{D}}$. Under this assumption, quantum gravity effects can be neglected. However the Horizon Quantum Mechanics approach advocates for significant black hole events even for $M_{\mathrm{B}}<\Lambda_{\mathcal{D}}$~\cite{Arsene:2016kvf,Gingrich:2020jpl}. 
	 
	In this work, we study the production of tensionless, non-rotating black hole at the LHC within the ADD framework. The semi-classical approach is used to describe black hole production, where the produced black hole undergo complete evaporation through Hawking radiation. In our analysis, we also take into account the effect of energy loss during black hole formation process (balding phase). This effect has been incorporated by means of the fractional energy loss parameter $\zeta \in [0,1)$, where $\zeta = 0$ corresponds to the case without any loss in the collision energy, whereas $\zeta > 0$ indicates that a fraction of the initial collision energy is radiated away before black hole reaches a stable stage. 
	It is to be noted that an alternative possibility is black hole remnants scenario, in which black hole can leave behind stable or long-lived remnants~\cite{Alberghi:2013hca}. These remnants can produce different experimental signatures compared to those expected from complete Hawking evaporation. The analysis is performed at a CM energy of $\sqrt{s} = 14$ TeV with an integrated luminosity of $349.4~\text{fb}^{-1}$. 
			
	The plan of the paper is as follows. In Section 2, we introduce the ADD model, which incorporates large extra spatial dimensions in which the fundamental Planck scale is lowered to the TeV range. In Section 3, we discuss the formation of black hole within the ADD framework at the LHC. We also examine black hole lifetimes to understand how quickly they decay after formation. Section 4 presents a numerical analysis of black hole production. We compute the production cross-sections of black hole and determine exclusion limits on black hole mass $M_{\mathrm{B}}$ for different values of $\Lambda_{\mathcal{D}}$, and $\mathcal{D}$. The analysis is performed for different values of loss parameter $\zeta$, allowing us to investigate its impact on black hole production rates and the resulting mass exclusion limits. Finally, we summarise our findings in Section 5.

\section{Black holes in ADD models}
	
	In the ADD model, $\mathcal{D}$ extra spatial dimensions are considered in addition to the usual three spatial dimensions. These $\mathcal{D}$ dimensions are compactified as closed loops with a compactification radius $\mathcal{R}_c$. In this framework, SM particles are confined to a four-dimensional spacetime called the brane which is embedded within a higher-dimensional spacetime, known as the bulk. Unlike SM particles, gravitons can spread throughout the bulk. This spreading of gravitational flux into the bulk effectively reduces the strength of gravity on the brane. For sufficient values of $\mathcal{R}_c$, the value of $M_{\mathrm{Pl}}$ can be lowered from its four-dimensional value of $M_{\mathrm{Pl}} \sim 10^{16}~\mathrm{TeV}$ to the value of only a few TeV~\cite{Meschini:2006gm}. In this case, gravity becomes very strong at high energies, which could lead to the production of black holes at the LHC~\cite{Banks:1999gd}. The relation between the four-dimensional Planck scale and the reduced gravity scale in extra spatial dimensions is given by the following relation~\cite{Arkani-Hamed:1998jmv}
	\begin{equation}
		M_{\mathrm{Pl}}^2 \sim \mathcal{R}_c^\mathcal{D} \, (\Lambda_{\mathcal{D}})^{2+\mathcal{D}},
	\end{equation}
	\noindent 
	where $M_{\mathrm{Pl}}$ is the four-dimensional Planck scale.	
	
	From the above expressions, it is obvious that for $\mathcal{D}$ dimensions, the values of $\mathcal{R}_c$ decreases as the $\Lambda_{\mathcal{D}}$ increases. For $\mathcal{D} = 1$, the value of $\mathcal{R}_c$ turned out to be extremely large (\ref{tab:Rc_meters}) at $\Lambda_{\mathcal{D}} = 10~\mathrm{TeV}$. Therefore, this scenario is ruled out because no deviation from Newtonian gravity has been observed experimentally so far~\cite{Long:2003ta}. For $\mathcal{D} = 2$ at $\Lambda_{\mathcal{D}} = 1~\mathrm{TeV}$, the value of $\mathcal{R}_c$ is approximately $0.24\,\mu\mathrm{m}$. This radius is large enough to be tested by gravity experiments~\cite{Landsberg:2004mj, Casse:2003pj}. Similarly for a fixed value of $\Lambda_{\mathcal{D}}$, the value of $\mathcal{R}_c$ decreases rapidly with increasing $\mathcal{D}$. For example, $\mathcal{D} = 3$ the value of $\mathcal{R}_c$ is of the order of a nanometer at $\Lambda_{\mathcal{D}} = 1~\mathrm{TeV}$, whereas for $\mathcal{D} = 7$, it decreases further to the femtometer scale. The detailed values of $\mathcal{R}c$ corresponding to different values of $\mathcal{D}$ and $\Lambda{\mathcal{D}}$ are shown in Table~\ref{tab:Rc_meters}.
	\begin{table}[ht]
		\centering
		\begin{tabular}{ccccc}
			\toprule
			\multirow{2}{*}{$D$} &
			\multicolumn{3}{c}{$R_c$ (m)}                                          \\
			\cmidrule(lr){2-4}
			& $\Lambda_D=1$ TeV & $\Lambda_D=10$ TeV & $\Lambda_D=100$ TeV         \\
			\midrule
			1 & $2.94\times10^{13}$ & $2.94\times10^{10}$ & $2.94\times10^{7}$     \\
			2 & $2.41\times10^{-3}$ & $2.41\times10^{-5}$ & $2.41\times10^{-7}$    \\
			3 & $1.05\times10^{-8}$ & $2.25\times10^{-10}$ & $4.85\times10^{-12}$  \\
			4 & $2.18\times10^{-11}$ & $6.89\times10^{-13}$ & $2.18\times10^{-14}$ \\
			5 & $5.37\times10^{-13}$ & $2.14\times10^{-14}$ & $8.50\times10^{-16}$ \\
			6 & $4.54\times10^{-14}$ & $2.11\times10^{-15}$ & $9.79\times10^{-17}$ \\
			7 & $7.78\times10^{-15}$ & $4.03\times10^{-16}$ & $2.09\times10^{-17}$ \\
			\bottomrule
		\end{tabular}\\
		\caption{Variation of compactification radius ($\mathcal{R}_c$) with respect to number of extra-dimenssions ($\mathcal{D}$) for $\Lambda_{\mathcal{D}} = 1, 10$ and $100~\mathrm{TeV}$.}
		\label{tab:Rc_meters}
	\end{table}
	
	The Newtonian gravitational constant $G$ may be written in the form of the Planck scale as $G = 1/M_{\mathrm{Pl}}^2$ in natural units. The gravitational potential between two test masses in $\mathcal{D}$ dimensions depends upon whether the separation $r$ is smaller or larger than the $\mathcal{R}_c$~\cite{Arkani-Hamed:1998sfv}, i.e. for $r > \mathcal{R}_c$, the gravitational potential is familiar to the four-dimensional Newtonian form and is given by, 
	\begin{equation}
		V(r)\sim \frac{m_1 m_2}{\Lambda_{\mathcal{D}}^{\mathcal{D}+2} \, \mathcal{R}_c^{\mathcal{D}}} \, \frac{1}{r}
			\sim \frac{m_1 m_2}{M_{\mathrm{Pl}}^{2}} \, \frac{1}{r}.
	\end{equation}
	\noindent 
	Whereas, for $r < \mathcal{R}_c$, it takes the following form~\cite{Cardoso:2005jq, Hoyle:2004cw},
	\begin{equation}
		V(r)\sim \frac{m_1 m_2}{\Lambda_{\mathcal{D}}^{\mathcal{D}+2}} \, \frac{1}{r^{\mathcal{D}+1}}.
	\end{equation}
	\noindent 
	
	\section{ADD Black Holes at LHC }
	
	At the LHC, black hole can be formed through the process $pp \to M_{\mathrm{B}} + X$, where `$X$' represents additional particles resulting from the collision~\cite {Cardoso:2005vb, Yoshino:2006dp}. The mass of the produced black hole is determined by the fraction of the initial CM energy that becomes gravitationally trapped within an event horizon during the collision process~\cite{Eardley:2002re}. Such black hole are formed due to the collision of a pair of highly relativistic partons emerging out of opposite moving protons, so that the gravitational field around the particles becomes very intense near the point where they almost collide and if impact parameter $b$ of the colliding partons is sufficiently small, their mutual gravitational attraction can concentrate enough energy within a compact region to form a black hole~\cite{Thorne:1972ji, Yoshino:2002tx}. In other words, the formation of black hole requires the squeezing of sufficient amount of collapsing energy into a region with circumference $ < 2\pi r_s$. This translates into $b$ shorter than $r_s$. According to Ref.~\cite{Dai:2007ki}, the critical value of $b$ required for black hole formation is given by
	\begin{equation}
		b < b_{\text{max}}
		= \frac{2~r_s}{\left[ 1 + \left( \frac{D-1}{2} \right)^2 \right]^{\frac{1}{D-2}}},
	\end{equation}
	\noindent	
	where $b_{\text{max}}$ represents the maximum allowed impact parameter for black hole formation. It determines the largest separation between the trajectories of the colliding particles for which gravitational collapse can still occur. The Schwarzschild radius ($r_s$) is given by~\cite{Dimopoulos:2001hw}
	\begin{equation}
		r_s = \frac{1}{\Lambda_{\mathcal{D}}} \left[ \frac{M_{\mathrm{B}}}{\Lambda_{\mathcal{D}}} \right]^{\frac{1}{\mathcal{D}+1}} f(\mathcal{D}),
		\label{eq:schwarzschild-radius}
	\end{equation}
	\noindent
	where $f(\mathcal{D})$ is a dimension dependent constant and is given by
	\begin{equation}
		f(\mathcal{D}) =\left(\frac{8 \pi^{-\frac{(\mathcal{D}+1)}{2}}\Gamma\left(\frac{\mathcal{D}+3}{2}\right)}{\mathcal{D}+2}\right)^{\frac{1}{\mathcal{D}+1}},
	\end{equation}
	\noindent
	where \(\Gamma(x)\) denotes the gamma function. The formation of a black hole can be described using a geometric cross-section approximation based on the impact parameter $b$, which corresponds to the effective horizon area of black hole formed in the collision. This is given by~\cite{Kanti:2004nr}
	\begin{equation}
		\sigma(p p \to M_{\mathrm{B}} + X) \approx \pi b_{\text{max}}^2 .
	\end{equation}
	\noindent
	If a black hole is produced at the LHC, it will have a non-zero Hawking temperature and will emits particles through Hawking radiation. The Hawking temperature is given by
		\begin{equation}
		T_H = \frac{\mathcal{D}+1}{4\pi r_s},
		\label{eq:hawking-temperature}
	\end{equation}
	\noindent
	where $T_H $ is the Hawking temperature. Since the Hawking temperature is inversely proportional to the $r_s$, smaller black hole have a relatively higher temperature and therefore evaporate more rapidly. As black hole radiates, its mass gradually decreases, which ultimately leads to an increase in its Hawking temperature as well as speeds up the evaporation process. 
	The lifetime of a black hole that emits Hawking radiation can be estimated as~\cite{Bleicher:2011uj}
	\begin{equation}
		\tau_{\mathrm{B}} \approx \frac{1}{\Lambda_{\mathcal{D}}} 
		\left( \frac{M_{\mathrm{B}}}{\Lambda_{\mathcal{D}}} \right)^{\frac{\mathcal{D}+3}{\mathcal{D}+1}},
	\end{equation}
	\noindent
	where $\tau_{\mathrm{B}}$ is the lifetime of black hole. The variation of $\tau_B$ with respect to various parameters is shown in Fig.~\ref{fig:lt_md_mb_dl2} which clearly indicate that for fixed ${\mathcal D}$, black hole lifetime becomes longer as $M_{\mathrm{B}}$ increases at fixed value of $\Lambda_{\mathcal{D}}$. However, for fixed ${\mathcal D}$ and $M_{\mathrm{B}}$, lifetime of black hole decreases rapidly as $\Lambda_{\mathcal{D}}$ increases.

	In this analysis, we assume the complete Hawking evaporation of black hole. We do not consider the possibility of stable or metastable remnants that might form if quantum gravity effects halt evaporation below a minimum mass, such as $M_{\mathrm{B}}\sim\Lambda_{\mathcal{D}}$ ~\cite{Alberghi:2013hca}.
	
 \section{Numerical Analysis }
 We now investigate the production of black holes at a CM energy of $\sqrt{s} = 14~\mathrm{TeV}$ with an integrated luminosity of $\int \mathcal{L}dt = 349.4~\mathrm{fb}^{-1}$. The analysis is performed within the framework of the ADD model. In this study, we focus only on black hole that are tensionless and non-rotating. In order to generate black hole events at the LHC, we use {\tt BlackMax}~\cite{ Dai:2009by}. The internal structure of the proton is described using the CTEQ6L parton distribution functions (PDF). The renormalization and factorization scales are taken to be equal to $\sqrt{\hat{s}}$, where $\sqrt{\hat{s}}$ is the partonic CM energy. During the formation phase, a black hole radiates away a fraction of its initial energy, linear and angular momentum before settling into a stationary state having mass $M_{\mathrm{B}}$, which corresponds to the minimal mass reached at the end of the balding phase~\cite{Hossenfelder:2004af}. These fractional losses are parameterized by three dimensionless coefficients $f_E$, $f_P$, and $f_L$~\cite{Dai:2007ki, Dai:2009by}, denoting the initial energy, linear momentum, and angular momentum, respectively. Since we consider non-rotating black hole in our analysis, only $f_E$ and $f_P$ loss parameters are relevant. The effects related to $f_L$ are neglected. Furthermore, we assume that energy and linear momentum losses occur proportionally, allowing both effects to be described by a single parameter $\zeta$ such that $f_E = f_P = \zeta$. The energy and linear momentum retained by black hole after the balding phase are given by
\begin{equation}
	E = E^{\mathrm{in}}(1-f_E), \qquad 
	P_z = P_z^{\mathrm{in}}(1-f_P), \qquad
\end{equation}
\noindent
where $E^{\mathrm{in}}$ and $P_z^{\mathrm{in}}$ represent the initial energy and linear momentum of black hole during the collision phase respectively, while $E$ and $P_z$ denote the energy and momentum remaining in black hole after it settles into a stationary state of mass $M_{\mathrm{B}}$. This stationary state is then considered as the initial state for the further evolution of black hole, including its evaporation through Hawking radiation. 

In this analysis, we consider the semi-classical (trans-Planckian) regime~\cite{Solodukhin:2002ui, Hsu:2002bd}, where black hole mass is sufficiently larger than reduced Planck scale. In this regime, quantum gravity effects are expected to be small and can therefore be neglected. The condition for the validity of the semi-classical approximation is given by
\begin{equation}
	\sqrt{\hat{s}} \ge M_{\mathrm{B}} > \Lambda_{\mathcal{D}}.
	\label{eq:mass-condition}
\end{equation}
We now proceed to study the production cross-sections of black hole in pp collisions. We investigate black hole production cross-section by varying $M_{\mathrm{B}}$ in the range of $2$ - $12~\mathrm{TeV}$ and $\Lambda_{\mathcal{D}}$ in the range of $1$ - $9~\mathrm{TeV}$. We consider dimensions $\mathcal{D}$ = $2$ - $7$ and different values of the loss parameter $\zeta$. To calculate the production cross-section, we consider two partons inside the incoming protons carrying momentum fractions $x_1$ and $x_2$ . For simplicity, we set the variables $v = x_1$ and $u = x_1 x_2$, which gives $x_2 = u/v$. The CM energy available for the partonic collision is therefore given by $\sqrt{\hat{s}} = \sqrt{u s}$. The total black hole production cross-section is obtained by integrating the parton-level cross-sections with the parton distribution functions over the allowed parton momentum fractions. Summing over all contributing partons, the total cross-section is given by
\begin{equation}
	\sigma_{pp \to \text{B} + X}(s; \mathcal{D}, \Lambda_{\mathcal{D}}) = 
	\int_{\frac{\Lambda_{\mathcal{D}}^2}{s}}^{1} \mathrm{d}u 
	\int_{u}^{1} \frac{\mathrm{d}v}{v} \, 
	\pi \left[b_{\max}\left(\sqrt{u s}; \mathcal{D}\right) \right]^2
	\sum_{i,j} f_i(v, Q^2) f_j\!\left(\frac{u}{v}, Q^2\right),
\end{equation}
\noindent
where $f_i(x)$ and $f_j(x)$ represent the PDFs for partons $i$ and $j$ evaluated at $Q^2 = \hat{s}$~\cite{Landsberg:2002sa, Kanti:2004nr}. 

We also note that there is an alternative approach known as the Horizon Quantum Mechanics (HQM) framework. In this approach, black hole production is still possible, but with a small probability, even for $M_{\mathrm{B}}<\Lambda_{\mathcal{D}}$. In this framework, the geometric cross-section is multiplied by a probability factor $P_{\mathrm{BH}}(E)$~\cite{Arsene:2016kvf}, which represents the probability of black hole formation. The probability of producing a black hole with mass $M_{\mathrm{B}}$ near $\Lambda_{\mathcal{D}}$ becomes strongly suppressed for higher $\mathcal{D}$ dimensions~\cite{Gingrich:2020jpl}.
 
 The exclusion limits presented in our work are obtained using the expected number of microscopic black hole events predicted by the ADD model. The 95\% CL exclusion limits on black hole mass are obtained from the expected signal yield. For each value of black hole mass, the expected number of observable signal events ($\mathcal{N}_{\text{B}}$) is calculated using,
 \begin{equation} 
 	\mathcal{N}_{\text{B} } = \sigma_{pp \to \text{B} + X}(s; \mathcal{D}, \Lambda_{\mathcal{D}})~C_{\text{eff}} \int \mathcal{L} dt,
 \end{equation}
 \noindent
 where $C_{\text{eff}}$ overall detection efficiency factor. In this analysis, we assume $C_{\text{eff}} = 0.6$, which corresponds to an event-selection efficiency in collider experiments.

\begin{figure}[H]
	\centering
	\includegraphics[width=\textwidth]{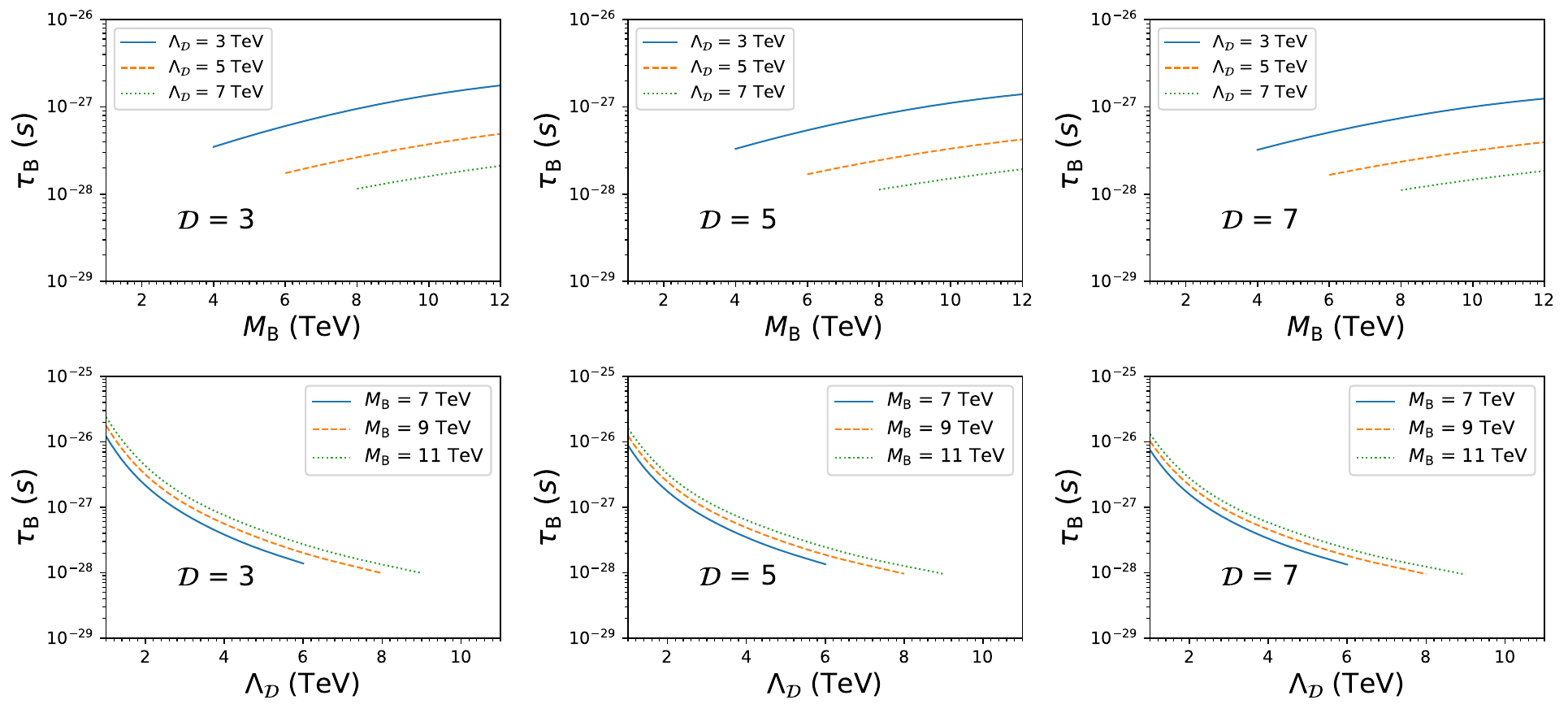} 
	\caption{Life-time of black hole vs $M_{\mathrm{B}}$ (top) and $\Lambda_{\mathcal{D}}$ (bottom) for $\zeta = 0$ and $\mathcal{D} = 3, 5,$ and $7$, respectively.}
	\label{fig:lt_md_mb_dl2}
\end{figure}

Furthermore, we study the lifetime ($\tau_{\mathrm{B}}$) of the formed black hole and examine their dependence on $M_{\mathrm{B}}$ and $\Lambda_{\mathcal{D}}$ across different $\mathcal{D}$ dimensions, as illustrated in Fig.~\ref{fig:lt_md_mb_dl2}. This analysis provides insight into black hole decay process. As a black hole radiates~\cite{Emparan:2000rs}, it gradually loses mass. As a results, its Schwarzschild radius also shrinks. Since Hawking temperature is inversely proportional to the $r_s$ (Eq.~\ref{eq:hawking-temperature}), black hole temperature rises as it loses mass, resulting in emission of increasingly high-energy particles.

The upper row shows that, for fixed values of $\Lambda_{\mathcal{D}}$, the lifetime of black hole increases rapidly as $M_{\mathrm{B}}$ increases for $\mathcal{D}=3$, $5$, and $7$. As black hole mass increases, the Hawking temperature decreases. Thus, heavier black hole evaporate more slowly and have longer lifetimes. The bottom row shows that as we increase $\Lambda_{\mathcal{D}}$, black hole lifetime decreases rapidly. At higher values of $\Lambda_{\mathcal{D}}$, the Schwarzschild radius for a given black hole mass is reduced, leading to a higher Hawking temperature and an accelerated evaporation process.

\begin{figure}[H]
	\centering
	\includegraphics[width=\textwidth]{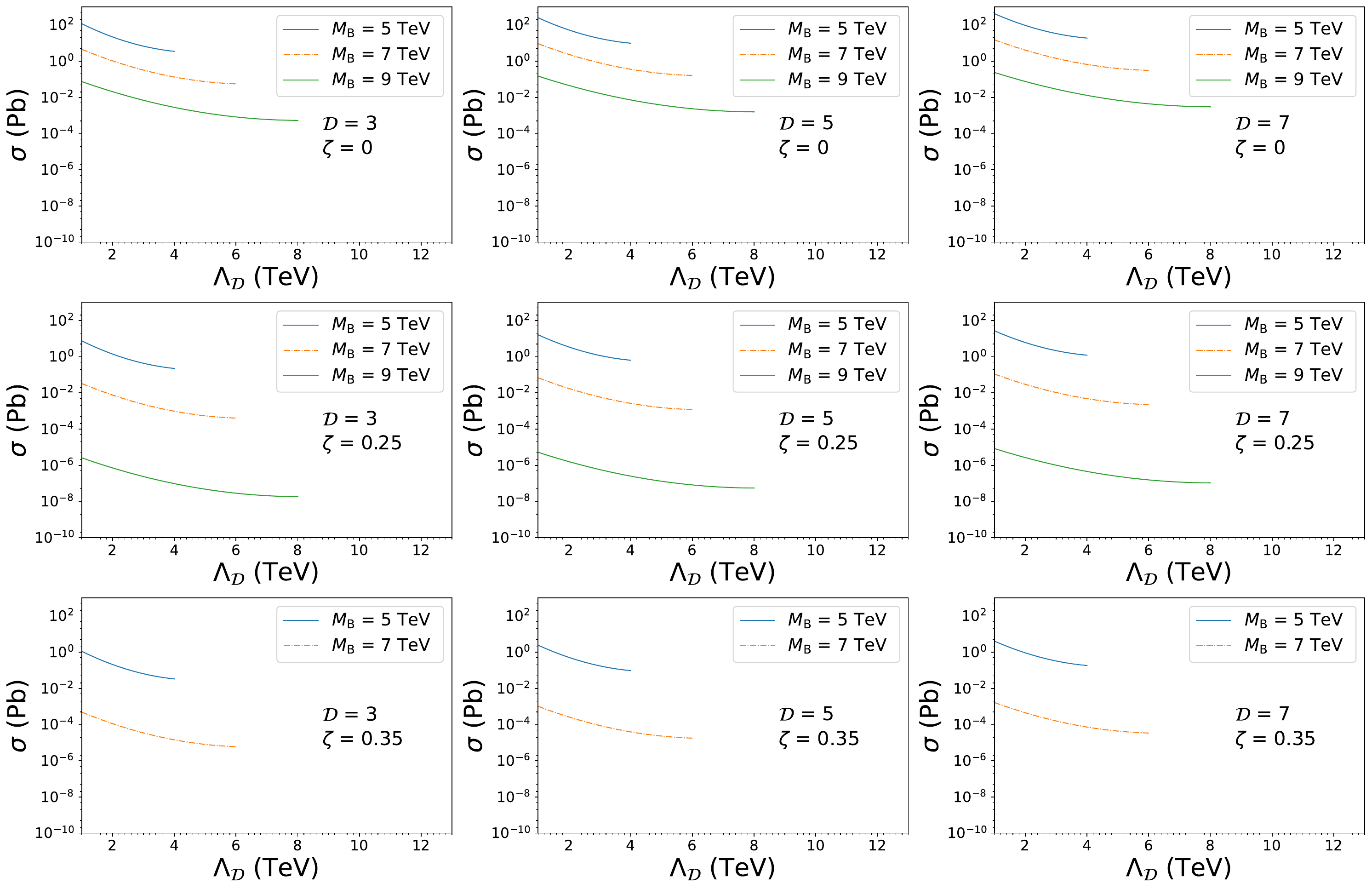} 
	\caption{Production cross-section versus $\Lambda_{\mathcal{D}}$ for dimensions $\mathcal{D} = 3$ (top), $5$ (middle), and $7$ (bottom) for $\zeta = 0$, $\zeta = 0.25$, and $\zeta = 0.35$.}		
	\label{fig:mb_x_md_d_l_2l}
\end{figure}

Next, we analyse the production cross-sections of black hole. The calculated cross-section depends significantly on parameters such as $M_{\mathrm{B}}$, $\Lambda_{\mathcal{D}}$, $\mathcal{D}$, and $\zeta$. Figure~\ref{fig:mb_x_md_d_l_2l} shows the variation of cross-section with $\Lambda_{\mathcal{D}}$ for $M_{\mathrm{B}} = 5$, $7$, and $9~\mathrm{TeV}$. It is observed that, for fixed values of $M_{\mathrm{B}}$, $\mathcal{D}$ and $\zeta$, cross-section decreases with increasing $\Lambda_{\mathcal{D}}$. This occurs because large values of $\Lambda_{\mathcal{D}}$ weaken the effective gravitational coupling and raise the kinematic threshold for black hole formation, requiring the colliding partons to carry a higher momentum fraction. Consequently, the PDFs drop rapidly, reducing the probability of finding partons with sufficient energy to form a black hole of mass $M_{\mathrm{B}}$.

\begin{figure}[H]
	\centering
	\includegraphics[width=\textwidth]{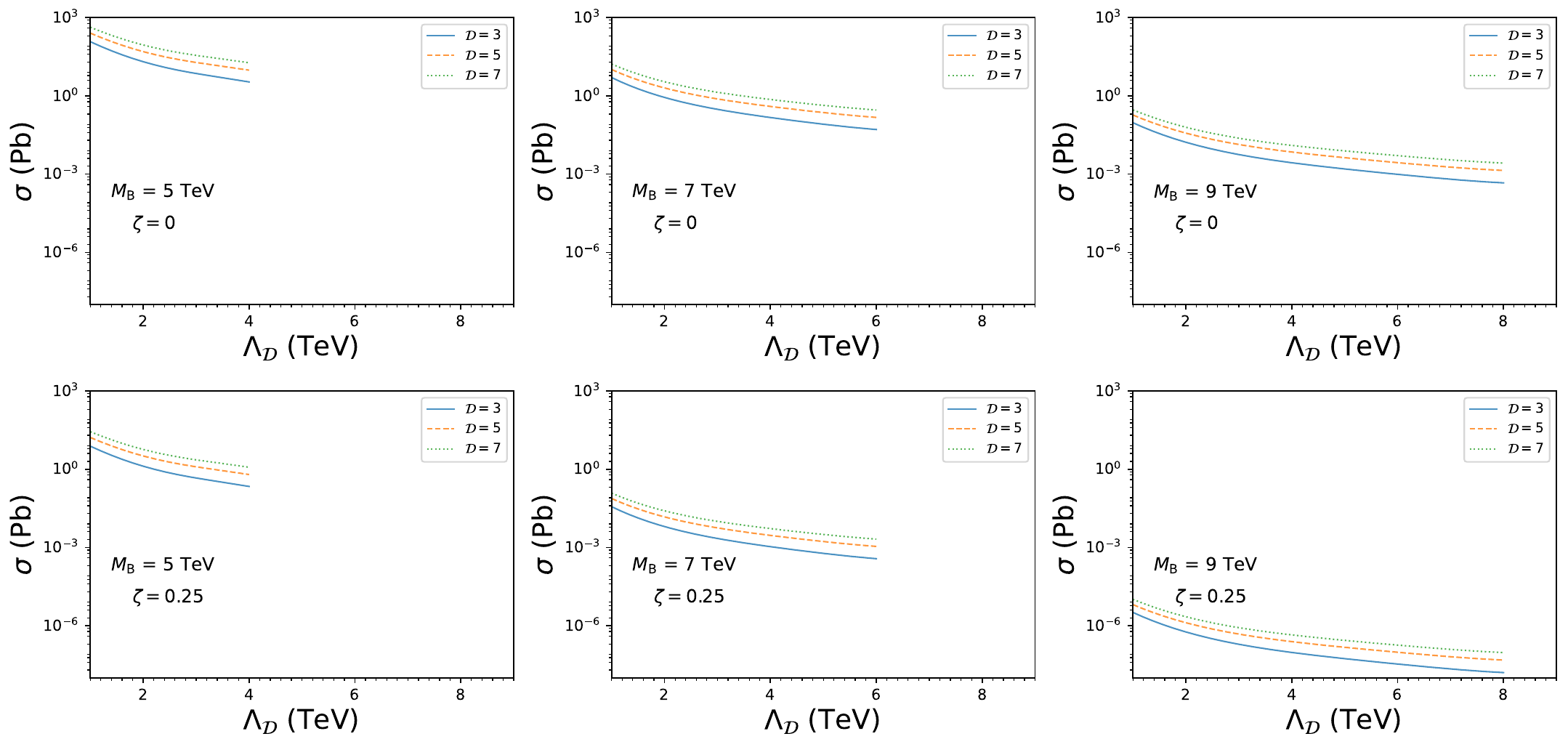} 
	\caption{Production cross-section versus $\Lambda_{\mathcal{D}}$ for $\mathcal{D} = 3, 5,$ and $7$, at values $M_{\mathrm{B}} = 5, 7, $ and $9~\mathrm{TeV}$ with $\zeta = 0$ (top) and $\zeta = 0.25$ (bottom).}
	\label{fig:mb_x_d_md2}
\end{figure}
 The impact of $\zeta$ is also evident in Fig.~\ref{fig:mb_x_md_d_l_2l} by comparing the different rows, where each row corresponds to a fixed value of $\zeta$. For $\zeta = 0~\rm and~0.25$, black hole with $M_{\mathrm{B}} = 9~\mathrm{TeV}$ is kinematically allowed, while as for $\zeta = 0.35$, black hole with mass $M_{\mathrm{B}} = 9~\mathrm{TeV}$ can not produced because the required partonic CM energy to create such a massive black hole is unavailable. This indicate that the required partonic CM energy exceeds the collider CM energy ($\sqrt{s} = 14~\mathrm{TeV}$). Therefore, the formation of such a heavy black holes is kinematically forbidden. We can also observe the effect of $\mathcal{D}$ in this figure. To see this more clearly, we plot the cross-section as a function of $\Lambda_{\mathcal{D}}$ in Fig.~\ref{fig:mb_x_d_md2}. For fixed value of $M_{\mathrm{B}}$ and $\zeta$, a slight increase in the cross-section is observed as $\mathcal{D}$ increases. This happens because the geometric cross-section increases as the $\mathcal{D}$ increases.

\begin{figure}[H]
	\centering
	\includegraphics[width=\linewidth]{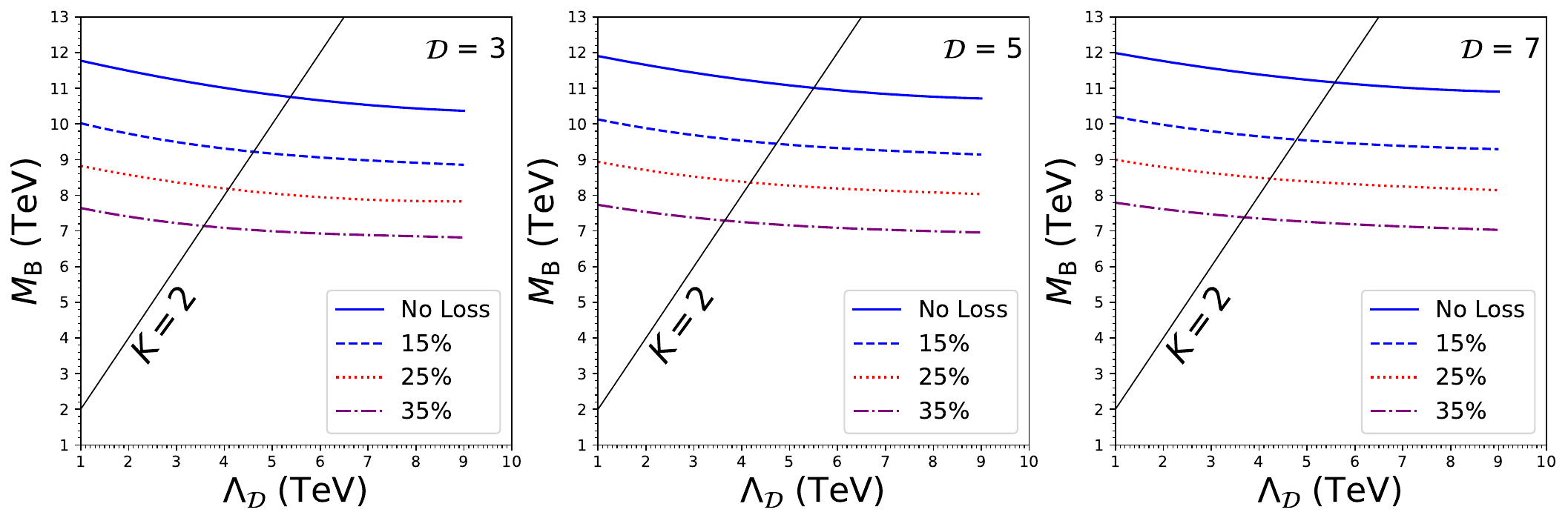}
	\caption{Exclusion curves in the $M_{\mathrm{B}}$ - $\Lambda_{\mathcal{D}}$ plane for $\mathcal{D} = 3$, $\mathcal{D} = 5$ and $\mathcal{D} = 7$ with different values of $\zeta$. The solid lines indicate constant ratios $K = M_{\mathrm{B}}/\Lambda_{\mathcal{D}}$.}
	\label{fig:md_mb_l_d_new2}
\end{figure}
Furthermore, we calculate the exclusion limits on black hole mass $M_{\mathrm{B}}$ across the various values of $\Lambda_{\mathcal{D}}$, $\mathcal{D}$, and $\zeta$. These exclusion limits are obtained for a CM energy $\sqrt{s}=14~\mathrm{TeV}$ and an integrated luminosity of $\int \mathcal{L}dt = 349.4~\mathrm{fb}^{-1}$ at a $95\%$ CL which corresponds to a significance of $1.96$ and therefore for observation of such events $\mathcal{N}_{\text B}$ must be at least $1.96 ~\simeq ~2$.

The obtained exclusion limits are summarized in Table~\ref{tab:bh_mass}. We further explore the accessible regions of the parameter space and identify the combinations of $\Lambda_{\mathcal{D}}$, $\zeta$, and $\mathcal{D}$ that satisfy the exclusion criteria. Figure~\ref{fig:md_mb_l_d_new2} shows the exclusion limits for $\mathcal{D} = 3$, $5$, and $7$, with each curve corresponding to a different value $\zeta = 0$, $0.15$, $0.25$, and $0.35$. The region below each curve is ruled out at the 95\% CL. In contrast, the region above each curve remains experimentally allowed. A reference line for the ratio $K = M_{\mathrm{B}}/\Lambda_{\mathcal{D}}$~\cite{ATLAS:2013bir} is shown in Fig.~\ref{fig:md_mb_l_d_new2} to indicate the semi-classical region. 

The semi-classical description of black hole formation is valid only when black hole mass is significantly larger than the reduced Planck scale (Eq.~\ref{eq:mass-condition}). This means that $r_s$ (Eq.~\ref{eq:schwarzschild-radius}), should be larger than the reduced Planck length, $\ell_{\mathcal{D}} \sim \Lambda_{\mathcal{D}}^{-1}$~\cite{Park:2012fe}. Therefore, the condition can be written as
\begin{equation}
	r_s \Lambda_{\mathcal{D}} > 1,
\end{equation}
\noindent
since $r_s$ depends on the ratio $ M_{\mathrm{B}}/\Lambda_{\mathcal{D}}$ (Eq.~\ref{eq:schwarzschild-radius}), this condition can also be written as
\begin{equation}
	K \equiv \frac{M_B}{\Lambda_D} > 1.
\end{equation}
 This is the minimum theoretical requirement for the semi-classical description. The condition $K \geq 2$ ensures that quantum gravity effects remain small and the semi-classical approximation remains reliable. This choice has been commonly used in previous studies and experimental searches~\cite{ATLAS:2013bir, ATLAS:2015aqb}. The line $K = 2$ is used only as a reference benchmark rather than a sharp physical boundary. Therefore, if the results are close to or below this value, quantum gravitational effects may become significant, and the semi-classical approximation may no longer be entirely reliable.

For the case of $\mathcal{D} = 7$ at $\Lambda_{\mathcal{D}} = 1~\mathrm{TeV}$, the analysis excluded black hole masses up to approximately $12.03~\mathrm{TeV}$ at the 95\% CL. When the loss parameter $\zeta$ increases, the exclusion boundaries shift toward lower values of $M_{\mathrm{B}}$. This shows that energy dissipation during the formation phase significantly shrink the exclusion parameter space. For example, when $\zeta = 0.35$ and $\Lambda_{\mathcal{D}} = 1~\mathrm{TeV}$, the excluded black hole mass is reduced to approximately $7.80~\mathrm{TeV}$ due to the energy loss effects.

\begin{table}[H] 
	\centering
	\begin{tabular}{cccccc}
		\toprule
		\multirow{2}{*}{\textbf{$\boldsymbol{\Lambda_{\mathcal{D}}}$ (TeV)}} & 
		\multirow{2}{*}{\textbf{$\boldsymbol{\mathcal{D}}$}} & 
		\multicolumn{4}{c}{\textbf{$\boldsymbol{M_{\mathrm{B}}}$ (TeV)}} \\ 
		\cmidrule(lr){3-6}
		& & $\boldsymbol{\zeta=0}$ 
		& $\boldsymbol{\zeta=0.15}$ 
		& $\boldsymbol{\zeta=0.25}$ 
		& $\boldsymbol{\zeta=0.35}$ \\
		\midrule
		
		\multirow{6}{*}{1} 
		& 2 & 11.74 & 9.96 & 8.76 & 7.58 \\
		& 3 & 11.83 & 10.03 & 8.84 & 7.65 \\
		& 4 & 11.90 & 10.10 & 8.90 & 7.70 \\
		& 5 & 11.95 & 10.14 & 8.95 & 7.74 \\
		& 6 & 11.99 & 10.18 & 8.98 & 7.77 \\
		& 7 & 12.03 & 10.21 & 9.01 & 7.80 \\
		\midrule
		
		\multirow{6}{*}{3} 
		& 2 & 11.01 & 9.33 & 8.21 & 7.10 \\
		& 3 & 11.19 & 9.49 & 8.35 & 7.22 \\
		& 4 & 11.31 & 9.60 & 8.45 & 7.31 \\
		& 5 & 11.40 & 9.68 & 8.52 & 7.37 \\
		& 6 & 11.47 & 9.73 & 8.57 & 7.42 \\
		& 7 & 11.53 & 9.79 & 8.62 & 7.46 \\
		\midrule
		
		\multirow{6}{*}{5} 
		& 2 & 10.59 & 8.97 & 7.90 & 6.83 \\
		& 3 & 10.83 & 9.18 & 8.08 & 6.99 \\
		& 4 & 10.98 & 9.32 & 8.20 & 7.10 \\
		& 5 & 11.09 & 9.44 & 8.29 & 7.17 \\
		& 6 & 11.18 & 9.49 & 8.36 & 7.22 \\
		& 7 & 11.25 & 9.55 & 8.41 & 7.27 \\
		\midrule
		
		\multirow{6}{*}{7} 
		& 2 & 10.29 & 8.68 & 7.71 & 6.69 \\
		& 3 & 10.57 & 8.97 & 7.92 & 6.88 \\
		& 4 & 10.74 & 9.12 & 8.02 & 6.96 \\
		& 5 & 10.87 & 9.24 & 8.11 & 7.03 \\
		& 6 & 10.96 & 9.30 & 8.17 & 7.08 \\
		& 7 & 11.04 & 9.33 & 8.22 & 7.12 \\
		\midrule
		
		\multirow{6}{*}{9} 
		& 2 & 10.03 & 8.54 & 7.60 & 6.62 \\
		& 3 & 10.33 & 8.86 & 7.83 & 6.82 \\
		& 4 & 10.53 & 9.00 & 7.94 & 6.89 \\
		& 5 & 10.68 & 9.15 & 8.04 & 6.96 \\
		& 6 & 10.79 & 9.23 & 8.10 & 7.00 \\
		& 7 & 10.88 & 9.29 & 8.15 & 7.03 \\
		
		\bottomrule
	\end{tabular}
	\caption{LHC exclusion limits $M_{\mathrm{B}}$ for values of $\Lambda_{\mathcal{D}}$, $\mathcal{D}$ and energy loss parameter $\zeta$ for $\sqrt{s} = 14~\mathrm{TeV}$ and $\int \mathcal{L} dt = 349.4~\mathrm{fb}^{-1}$ at $95\%$ CL}
	\label{tab:bh_mass}
\end{table} 

\section{Results and Discussion}
\label{summary}

We investigated the production of microscopic black hole arising due to large extra spatial dimensions in the context of the ADD model at the LHC with $\sqrt{s} = 14$ TeV and an integrated luminosity of $\int {\mathcal L} dt = 349.4$ fb$^{-1}$ for the additional dimensions ranging up to 7. Throughout our investigation, we assume that black hole are non-rotating and tensionless. The dependence of the production cross-sections at LHC have been presented as Figs.~2 - 3 for the reduced Planck scale, $\Lambda_{\mathcal{D}} \in [1, 9 ]~\rm TeV$, $M_{B} \in [2, 12 ]~\rm TeV$, and extra spatial dimensions $\mathcal{D} \in [2, 7]$. Our study also considers the energy loss that occurs during black hole formation phase. This energy loss is described by the parameter $\zeta \in [0, 1)$, where $\zeta = 0$ means no loss and $\zeta > 0$ represents the fractional energy loss.

From the Figs. 2 - 3, it is clear that the production cross-section decreases significantly with increasing values of $\Lambda_{\mathcal{D}}$ and $\zeta$, reflecting a reduction in the effective energy 
available for the formation of a black hole of mass $M_{\mathrm{B}}$. The cross-section increases with the $\mathcal{D}$ due to the rise of the Schwarzchild radius. We also presented the LHC exclusion plots for $\sqrt{s} = 14$ TeV and $\int {\mathcal L} dt = 349.4$ fb$^{-1}$ for the aforementioned ranges of the parameter space as Figs.~4 for different values of ${\mathcal D}$ and $\zeta$.

Our analysis reveals that for $\zeta = 0$, black hole with $M_{\mathrm{B}} \leq 11.83~\mathrm{TeV}$ are disfavored in the case of dimensions $\mathcal{D} = 3$, for a 
$\Lambda_{\mathcal{D}} \sim 1~\mathrm{TeV}$. This exclusion limit decreases to approximately $10.33~\mathrm{TeV}$ when the $\Lambda_{\mathcal{D}} = 9~\mathrm{TeV}$ (Table~\ref*{tab:bh_mass}), reflecting the suppression of the production cross-section at higher values of the $\Lambda_{\mathcal{D}}$. A significant reduction is also observed when loss effects are included. For $\zeta = 0.35$, the excluded black hole mass reduces to $7.65~(6.82)~\mathrm{TeV}$ for $\Lambda_{\mathcal{D}} = 1~(9)~\mathrm{TeV}$. This significant decrease highlights the energy dissipation during black hole formation process, which effectively reduces the available energy for black hole production. We further extend the analysis for $\mathcal{D}=7$, the exclusion limits are substantially modified due to the geometric effects associated with dimensions.

 In this case, for $\zeta = 0$, black holes with masses up to $12.03~(10.88)~\mathrm{TeV}$ 
are excluded for $\Lambda_{\mathcal{D}} = 1~(9)~\mathrm{TeV}$ and for $\zeta = 0.35$, the corresponding limits reduce to $7.80~(7.03)~\mathrm{TeV}$ at the same values of $\Lambda_{\mathcal{D}}$. This clearly reflect that the allowed region suppresses as we move to higher values of $\zeta$.


\end{document}